\documentclass[sn-mathphys, Numbered]{sn-jnl}

\usepackage{graphicx}%
\usepackage{multirow}%
\usepackage{amsmath,amssymb,amsfonts}%
\usepackage{amsthm}%
\usepackage{mathrsfs}%
\usepackage[title]{appendix}%
\usepackage{xcolor}%
\usepackage{textcomp}%
\usepackage{manyfoot}%
\usepackage{booktabs}%
\usepackage{algorithm}%
\usepackage{algorithmicx}%
\usepackage{algpseudocode}%
\usepackage{listings}%

\usepackage{physics}
\usepackage{geometry}
\begin{document}

\title[Article Title]{$ \Delta $ baryon spectroscopy in lattice QCD}


\author*[1]{\fnm{Liam} \sur{Hockley}}\email{liam.hockley@adelaide.edu.au} 

\author[1]{\fnm{Waseem} \sur{Kamleh}}

\author[1]{\fnm{Derek} \sur{Leinweber}}

\author[1]{\fnm{Anthony} \sur{Thomas}}

\affil[1]{\orgdiv{ARC Special Research Centre for the Subatomic Structure of Matter (CSSM)}, \orgname{Department of Physics, The University of Adelaide}, \orgaddress{\state{SA}, \postcode{5005}, \country{Australia}}}

\abstract{A variational analysis is performed within the framework of lattice QCD to extract the masses of the spin-3/2 positive parity $ \Delta^+ $ baryons, including radial excitations. 2+1 flavour dynamical gauge-field configurations provided by the PACS-CS collaboration via the ILDG are considered. To improve our interpolator basis, we perform an iterative process of source and sink smearing and solve a generalised eigenvalue problem using the resulting fermion operators. We obtain a clear signal for the ground and first excited states at a light quark mass corresponding to $ m_\pi = 413 $ MeV. Furthermore, we show that one can use the eigenvectors obtained in this method to investigate the nature of these states, allowing us to classify our results as $ 1s $ and $ 2s $ states for the ground and first excited states respectively. Finally, we briefly highlight the method of Hamiltonian Effective Field Theory which can be used to make comparison with quark model expectations.}




\maketitle

\section{Introduction}\label{sec1}
The strong nuclear force underpins much of the world around us, giving rise to the hadronic matter in our universe through the interaction of quarks and gluons. While the equations governing this fundamental force of nature are captured by the theory of Quantum Chromodynamics (QCD), a complete understanding is elusive due to the non-perturbative nature of QCD in the low energy regime. This is precisely the realm of hadrons such as the nucleon, which is so fundamental to our understanding of the universe.

Fortunately, lattice QCD is a technique designed specifically for tackling the non-perturbative part of QCD and has seen great advances in its application alongside developments in supercomputing technology. By discretising space-time and employing a number of systematically improvable approximations, the once impossible task of computing the mass of low-lying baryons is now tractable at percent-level precisions.

The nucleon spectrum has historically presented an interesting problem for the community, focused on the $ N^*(1440) $ Roper resonance \cite{Burkert:2017djo, Wu:2017qve}. The Roper has been of particular interest in the past due to its stark juxtaposition with simple theory expectations; while it is the second state in the nucleon spectrum and has positive parity, a simple quark model without fine tuning predicts the second state should have negative parity. The closely related $ \Delta $ baryons also exhibit this anomalous energy level ordering and present a similar issue for our understanding of the strong force.

One way of understanding the Roper and other low-lying baryon resonances such as the $ \Delta $ has developed in recent years through Hamiltonian Effective Field Theory (HEFT) \cite{Hall:2013qba, Hall:2014uca, Liu:2015ktc, Liu:2016uzk, Liu:2016wxq, Wu:2017qve, Abell:2021awi}. Traditionally, this approach fits experimental data and brings this information to the finite volume of the lattice in the form of predictions of the finite volume spectra. Thus one main aim of this paper is to provide results for the $ \Delta $ baryon spectrum which can be confronted with future HEFT analyses.

Another aim for this work is to present the lattice techniques of spin and parity projection coupled with a variational analysis to isolate states in the spectrum. These techniques have been developed separately in different contexts \cite{Mahbub:2013ala, Zanotti:2003fx} and we aim to show how together they allow one to isolate and classify states in, for example, the $ J^P = 3/2^+,\ \Delta^+ $ spectrum.

\section{Energies from Correlation Functions}\label{sec2}
In lattice QCD we calculate correlation functions by first calculating the relevant quark propagators. At the baryon level, these correlation functions are defined as
\begin{equation}
	\mathcal{G}^{ij}_{\mu\nu}(t,\boldsymbol{p}) = \sum_{\boldsymbol{x}} e^{-i\boldsymbol{p}\cdot \boldsymbol{x}} \mel{\Omega}{T\{ \chi^i_\mu(x) \bar{\chi}^j_\nu(0) \}}{\Omega}\,.
\end{equation}
The operator $ \bar{\chi}^j_\nu(0) $ acts on the vacuum to create states at the space-time point 0. These states then propagate through Euclidean time $ t $ before being annihilated at a new space-time point $ x = (t,\boldsymbol{x}) $ by the operator $ \chi^i_\mu(x) $. $ T $ indicates time ordering of the operators.

If we are interested in the excited states of hadrons, we choose a set of operators with some non-zero overlap with the states of interest. For baryons, we have the complete set of states
\begin{equation}
	\sum_{B,\boldsymbol{p}\,',s} \ket{B,\boldsymbol{p}\,',s}\bra{B,\boldsymbol{p}\,',s} = I\,,
\end{equation}
where $ B $ labels different baryons with momenta $ \boldsymbol{p}' $ and spins $ s $. This allows us to obtain
\begin{align}
	\mathcal{G}^{ij}_{\mu\nu}(t,\boldsymbol{p}) 
	&= \sum_{B,\boldsymbol{p}\,',s} \sum_{\boldsymbol{x}} e^{-i\boldsymbol{p}\cdot \boldsymbol{x}} \mel{\Omega}{\chi^i_\mu(x)} {B,\boldsymbol{p}\,',s}\mel{B,\boldsymbol{p}\,',s} {\bar{\chi}^j_\nu(0)}{\Omega}\,.
\end{align}
We then use the translational operator to rewrite $ \chi^i_\mu(x) $ as
\begin{equation}
	\chi^i_\mu(x) = e^{Ht}e^{-i\boldsymbol{P}\cdot\boldsymbol{x}}\chi^i_\mu(0)e^{i\boldsymbol{P}\cdot\boldsymbol{x}}e^{-Ht}\,.
\end{equation}
This allows us to simplify the expression for the correlation function to
\begin{equation}
	\mathcal{G}^{ij}_{\mu\nu}(t,\boldsymbol{p}) 
	= \sum_{B,s} 
	e^{-E_Bt} \mel{\Omega}{\chi^i_\mu(0)} {B,\boldsymbol{p},s} \mel{B,\boldsymbol{p},s}{\bar{\chi}^j_\nu(0)}{\Omega}\,. \label{correlator}
\end{equation}

Now we set the momentum to $ \boldsymbol{p}=0 $, so that $ E_{B} = M_{B} $. We also introduce the parity projection operators
\begin{equation}
	\Gamma^\pm = \frac{1}{2} \qty(\gamma_0 \pm \mathbb{I})
\end{equation}
so one can then compute the parity projected correlator and sum over the spatial Lorentz indices $ \mu = \nu = n $ \cite{Leinweber:1992hy}
\begin{equation}
	G^\pm_{ij}(t,\boldsymbol{0}) \equiv \tr_{sp} \qty[\Gamma^\pm \sum_n \mathcal{G}^{ij}_{nn}(t,\boldsymbol{0})]\,. 
\end{equation}

By replacing the matrix elements in Eq.~\eqref{correlator} with appropriate functions (i.e. Rarita-Schwinger spinors for the case of spin-3/2 baryons, or regular Dirac spinors for spin-1/2), one can show \cite{Leinweber:1992hy} that the result is a series of decaying exponentials governed by the mass of the baryon states
\begin{equation}
	G^\pm_{ij}(t,\boldsymbol{0}) = \sum_{B^\pm} \lambda_{iB^\pm} \overline{\lambda}_{jB^\pm} e^{-M_{B^\pm}t}\,, \label{trace_corr}
\end{equation}
where $ \lambda_{i B^\pm} $ and $ \overline{\lambda}_{j B^\pm} $ are coupling strengths between the interpolating fields $ \chi^i_\mu $ and $ \overline{\chi}^j_\nu $ and the parity projected baryon states $ B^{\pm} $. We note that these coupling strengths can be taken to be real by considering both the original gauge-field links and their complex conjugates, weighted equally in the ensemble average \cite{Leinweber:1992hy, Mahbub:2013ala}.

In order to extract the ground state mass of a particular baryon state, one takes the long-time limit in which all the excited states have decayed off. Explicitly,
\begin{equation}
	G^\pm_{ij}(t,\boldsymbol{0}) \overset{t\to\infty}{=} \lambda^\pm_{i0}\overline{\lambda}^\pm_{j0} e^{-M_{0^\pm}t}\,, \label{groundcorr}
\end{equation}
where the $ \lambda^\pm_{i0} $ and $\overline{\lambda}^\pm_{j0} $ are couplings of baryon interpolators to the lowest lying state. With this preliminary discussion out of the way, we now discuss a number of techniques used to extract the various states in the $ \Delta $ spectrum.

\section{Spin Projection}\label{sec3}
We use a standard interpolating field for the $ \Delta^+ $ given by
\begin{equation}
	\chi^{\Delta^+}_\mu (x) = \frac{1}{\sqrt{3}} \epsilon^{abc} \qty[ 2\qty( u^{Ta}(x) C\gamma_\mu d^b(x) ) u^c(x) + \qty( u^{Ta}(x) C \gamma_\mu u^b(x) ) d^c(x) ]\,. \label{deltainterp}
\end{equation}
This operator has overlap with both spin-1/2 and spin-3/2 states so we need to perform spin projection to guarantee we extract the masses of the desired states. The spin-1/2 and spin-3/2 projection operators are given by \cite{Benmerrouche:1989uc, Zanotti:2003fx}

\begin{equation}
	P^{3/2}_{\mu\nu} (p) = g_{\mu\nu} - \frac{1}{3}\gamma_\mu\gamma_\nu - \frac{1}{3p^2}(\gamma\cdot p\ \gamma_\mu p_\nu + p_\mu \gamma_\nu\ \gamma\cdot p)\,, \label{proj32}
\end{equation}
\begin{equation}
	P^{1/2}_{\mu\nu} (p) = g_{\mu\nu} - P^{3/2}_{\mu\nu}(p)\,. \label{proj12}
\end{equation}

Although Eq.~(\ref{proj32}) looks somewhat involved and cumbersome, we can do a few things to simplify these operators. First, in our lattice calculations the mass is on shell, so we have
\begin{equation}
	p = (E,\boldsymbol{p}) = (\sqrt{\boldsymbol{p}^2 + m^2},\boldsymbol{p})\,.
\end{equation}

We'll consider the particles being at rest, $ \boldsymbol{p} = 0 $, in which case we get $ p_0 = m $ and as usual $ p^2 = m^2 $. Further, we also have
\begin{align}
	\gamma\cdot p &= \gamma^0p^0 - \boldsymbol{\gamma}\cdot \boldsymbol{p} = \gamma^0m\,, \\
	p_\mu &= m\delta_{\mu 0} = mg_{\mu 0}\,.
\end{align}

With these results in mind, we simplify our projection operator to
\begin{align}
	P^{3/2}_{\mu\nu} (\boldsymbol{p}=0)
	&= g_{\mu\nu} - \frac{1}{3}\gamma_\mu\gamma_\nu - \frac{1}{3}(\gamma_0 \gamma_\mu g_{\nu 0} + g_{\mu 0} \gamma_\nu \gamma_0)\,. \label{reduced}
\end{align}

One can then show by using the properties of the $ \gamma $-matrices and the metric, that the elements of the projectors obey
\begin{equation}
	P^{3/2}_{0 0} (\boldsymbol{p}=0) = 
	P^{3/2}_{0 n} (\boldsymbol{p}=0) = 
	P^{3/2}_{m 0} (\boldsymbol{p}=0) = 0\,, \qquad
	P^{3/2}_{m n} (\boldsymbol{p}=0) = g_{m n} - \frac{1}{3}\gamma_m\gamma_n\,. \label{res1}
\end{equation}
where $ m,\, n $ are spatial Lorentz indices.

We can immediately get the corresponding results for the spin-1/2 projection operator by making use of Eq.~(\ref{proj12}) and (\ref{res1}) and we obtain
\begin{equation}
	P^{1/2}_{0 0} (\boldsymbol{p}=0) = \mathbb{I}\,, \qquad 
	P^{1/2}_{0 n} (\boldsymbol{p}=0) =
	P^{1/2}_{m 0} (\boldsymbol{p}=0) = 0\,, \qquad
	P^{1/2}_{m n} (\boldsymbol{p}=0) = \frac{1}{3}\gamma_m\gamma_n\,. \label{res2}
\end{equation}

With the spin projection operators in hand, a spin-$ s $ projected correlation function is then given by
\begin{equation}
	\mathcal{G}^s_{\mu\nu} = \sum_{\sigma,\lambda = 1}^{4} \mathcal{G}_{\mu\sigma} g^{\sigma\lambda} P^s_{\lambda\nu}\,.
\end{equation}

This spin projection is performed prior to the parity projection and trace in Eq.~\eqref{trace_corr}. Thus we can obtain results for the masses of both the spin-1/2 and spin-3/2 $ \Delta^+ $ states without needing to generate the correlation functions more than once, saving on overall compute time.

\section{Source and Sink Smearing}\label{sec4}
In order to improve the overlap of the interpolating fields with the states of interest, we apply Gaussian smearing to the spatial components of the interpolating fields. The general procedure is to take some fermion field $ \psi_i(t, \boldsymbol{x}) $ and iteratively apply a smearing function $ F(\boldsymbol{x},\boldsymbol{x}') $. Explicitly, this takes the form
\begin{equation}
	\psi_i(t,\boldsymbol{x}) = \sum_{\boldsymbol{x}'}\, F(\boldsymbol{x},\boldsymbol{x}')\, \psi_{i-1}(t,\boldsymbol{x}')
\end{equation}
where the smearing function is
\begin{equation}
	F(\boldsymbol{x},\boldsymbol{x}') = (1-\alpha)\delta_{x,x'} 
	+ \frac{\alpha}{6} \sum_{\mu = 1}^{3} \qty[U_\mu(x)\delta_{x',x+\hat{\mu}} + U^\dagger_\mu(x - \hat{\mu})\delta_{x',x-\hat{\mu}}]\,.
\end{equation}
We take the smearing parameter to be $ \alpha=0.7 $ in our calculations. The use of repeated applications of the smearing function controls the width of our source. 

\begin{figure}[h]
	\centering
	\includegraphics[width=0.4\linewidth]{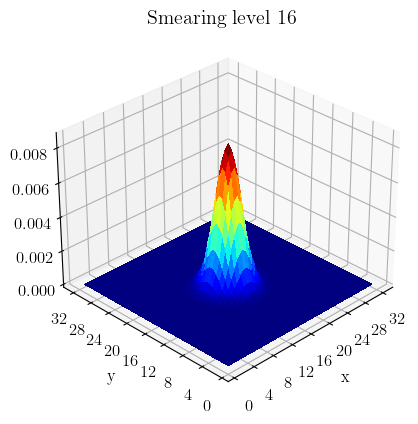}
	\includegraphics[width=0.4\linewidth]{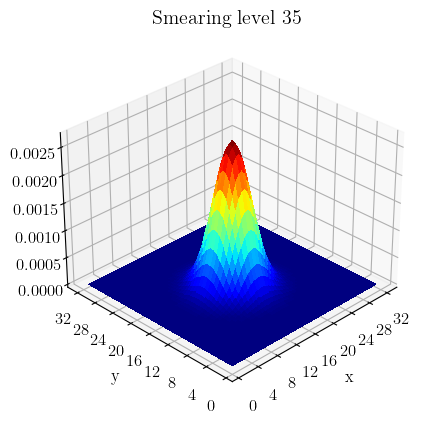}
	\includegraphics[width=0.4\linewidth]{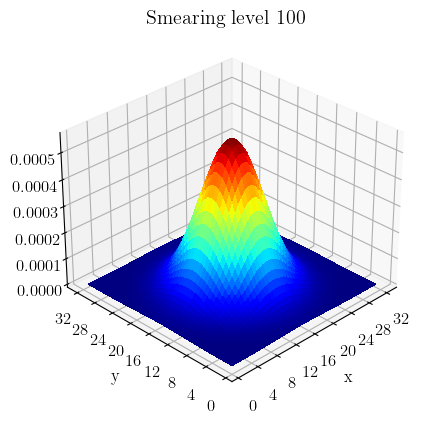}
	\includegraphics[width=0.4\linewidth]{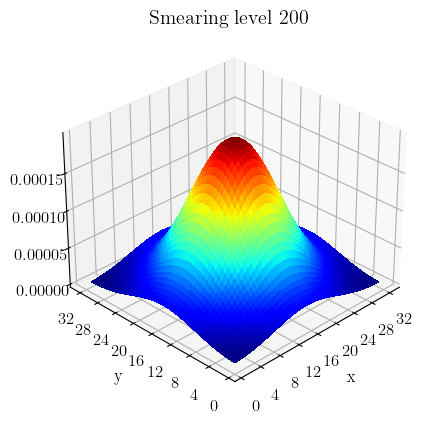}
	\caption{ Gaussian distributions resulting from smearing a point source with 16, 35, 100 or 200 sweeps of smearing. Note the progressively wider Gaussians obtained from repeating the smearing.}\label{fig1}
\end{figure}

Smearing allows us to more accurately represent a fermion bound within a hadron by smearing out a point source/sink so it achieves some finite width. This is demonstrated in Figure~(\ref{fig1}) where higher levels of smearing lead to Gaussian distributions of larger RMS radius. This broadening of the sources and sinks leads to an improved overlap of the resulting interpolating field with the states in the spectrum. Additionally, a combination of smeared sources and sinks can also give an indication of radial excitations in the resulting spectrum. This will be discussed further in Section~\ref{sec8}.

\section{Variational Analysis}\label{sec5}
We now resume our main discussion of how one extracts the masses of a baryon spectrum. Recall that in Section~\ref{sec2} we showed how one can readily obtain the mass of baryon ground states by simply taking the leading order contribution to the correlation function as in Eq.~\eqref{groundcorr}. In order to isolate states of higher energies, a more nuanced approach is required, since these states are at sub-leading order in the exponential series. We make use of the well documented variational analysis method \cite{Michael:1985ne, Mahbub:2013ala} in order to extract the ground state mass, as well as the 1st excited state mass. We also obtain results for the 2nd and 3rd excited state masses, and will report those in a forthcoming publication.

To extract $ N $ states, we use $ N $ interpolating fields (one for each state). We generate our interpolating fields by using the same base field for the $ \Delta^+ $ baryon, and applying different levels of smearing. These smearing levels are chosen for their coupling to the states of interest, and the use of smearing on both sources and sinks allows us to construct a matrix of correlation functions. This correlation matrix is written as

\begin{equation}
	G^\pm_{ij}(t) = \sum_{\alpha=0}^{N-1}\, \lambda^\alpha_i \overline{\lambda}^\alpha_j e^{-m_\alpha t}\,. \label{corr3}
\end{equation}

Here the $ \lambda^\alpha_i $ and $ \overline{\lambda}^\alpha_j $ are essentially the same as seen before in Section~\ref{sec2}, though we now use the $ \alpha $ index to distinguish energy states. In other words, they are couplings of the interpolators $ \chi^i_\mu $ and $ \overline{\chi}^j_\nu $ at the source and sink to the various energy eigenstates $ \alpha = 0, \dots , N-1 $. Finally, $ m_\alpha $ is the mass of the state $ \alpha $.

From here, we now aim to construct linear combinations of our operators to cleanly isolate the $ N $ states in the baryon spectrum. Labelling these baryon states $ \ket{B_\alpha} $, we thus wish to construct the superpositions
\begin{align}
	\overline{\phi}^\alpha_\mu 	&= \sum_{i = 1}^{N} u^\alpha_i \overline{\chi}^i_\mu\,, \\
	\phi^\alpha_\mu &= \sum_{i = 1}^{N} v^\alpha_i \chi^i_\mu\,, \label{phibar}
\end{align}
such that
\begin{align}
	\mel*{B_\beta, p, s}{\overline{\phi}^\alpha_\mu}{\Omega} &= \delta_{\alpha\beta} \overline{z}^\alpha \overline{u}_\mu(\alpha,p,s)\,, \\
	\mel*{\Omega}{\phi^\alpha_\nu}{B_\beta, p, s} &= \delta_{\alpha\beta} z^\alpha u_\nu(\alpha,p,s)\,, \label{phi}
\end{align}
where $ u_\mu(\alpha, p, s) $ is a Rarita-Schwinger spin vector. Here, the $ z^\alpha $ and $ \overline{z}^\alpha $ are the couplings of the superpositions $ \phi^\alpha_\mu $ and $ \overline{\phi}^\alpha_\nu $ to the state $ \ket{B_\alpha} $. The $ u^\alpha_i $ and $ v^\alpha_i $ are simply the weights for the superposition of fields, using the basis of smeared interpolating fields.

At this point, we will attempt to construct an eigenvalue problem to solve for both $ \boldsymbol{u}^\alpha $ and $ \boldsymbol{v}^\alpha $. Noting that since $ G_{ij}(t) = G_{ji}(t) $ in the ensemble average, we introduce an improved unbiased estimator of the correlation matrix $ 1/2\qty[G_{ij}(t) + G_{ji}(t)] $. This provides us with a correlation matrix which is symmetric, so we can simultaneously compute $ \boldsymbol{u}^\alpha $ and $ \boldsymbol{v}^\alpha $ as discussed below.

Multiplying Eq.~\eqref{corr3} on the right by $ u^\alpha_j $ we obtain
\begin{equation}
	G^\pm_{ij}(t) u^\alpha_j = \lambda^\alpha_i \overline{z}^\alpha e^{-m_\alpha t}\,.
\end{equation}

Then, since the exponential is the only time dependent part of the correlation function, we can form a  recurrence relation at some time after source insertion by introducing the variational parameters $ t_0 $ and $ \Delta t $:
\begin{equation}
	G_{ij}(t_0 + \Delta t) u^\alpha_j = e^{-m_\alpha \Delta t} G_{ij}(t_0) u^\alpha_j\,.
\end{equation}

Then, multiplying on the left by the inverse $ [G_{ij}(t_0)]^{-1} $ and suppressing the indices $ i $ and $ j $ gives 
\begin{equation}
	[G(t_0)^{-1} G(t_0 + \Delta t)] \boldsymbol{u}^\alpha = e^{-m_\alpha \Delta t}\boldsymbol{u}^\alpha\,, \label{genu}
\end{equation}
which we recognise as an eigenvalue equation for the vector in interpolator space $ \boldsymbol{u}^\alpha $.

Similarly, by premultiplying Eq.~\eqref{corr3} by $ v^\alpha_i $ we get
\begin{equation}
	v^\alpha_i G_{ij}(t_0 + \Delta t) = e^{-m_\alpha \Delta t} v^\alpha_i G_{ij}(t_0)
\end{equation}
from which we can arrive at our second eigenvalue equation (this time for $ \boldsymbol{v}^\alpha $):
\begin{equation}
	\boldsymbol{v}^\alpha G(t_0 + \Delta t)[G(t_0)]^{-1} = e^{-m_\alpha \Delta t} \boldsymbol{v}^\alpha\,. \label{genv}
\end{equation}

Both Eq.~\eqref{genu} and \eqref{genv} need to be solved simultaneously for each given pair of variational parameters, and we do so using a generalised eigenvalue problem solver. Solving for these eigenvectors automatically gives us the weights for the superpositions of interpolating fields, by construction. 

Finally, the eigenstate and parity projected correlation function is then taken to be
\begin{equation}
	G^\alpha_\pm \equiv v^\alpha_i G^\pm_{ij}(t) u^\alpha_j\,.
\end{equation}
We can clearly see that these eigenvectors have isolated a single state in the baryon spectrum, exactly as we set out to do. We then construct the effective mass function
\begin{equation}
	M^\alpha_{\text{eff}}(t) = \frac{1}{\delta t} \ln(\frac{G^\alpha_\pm(t,\boldsymbol{0})}{G^\alpha_\pm(t+\delta t,\boldsymbol{0})})\,. \label{effmass}
\end{equation}
As a note, $ \delta t $ should not be confused with $ \Delta t $. The latter is a variational parameter which we set to allow us to get a few time slices away from the source time. The former is typically taken to be small and is set independently of the variational parameters. We take $ \delta t = 2 $ in our calculations.

It is also worth noting that the eigenvectors $ \boldsymbol{u}^\alpha $ and $ \boldsymbol{v}^\alpha $ must also be equal since $ G_{ij}(t) $ is a real symmetric matrix. From here on, we will refer to the $ \boldsymbol{u}^\alpha $ vector, for simplicity.

The effective mass defined in Eq.~\eqref{effmass} can be computed for various discrete values of $ t $ and then plotted as a function of time. As usual, one looks for time intervals over which the effective mass plot plateaus, indicating that all contamination from unwanted states has decayed away in the exponential series. We finally perform a covariance matrix analysis \cite{Mahbub:2013bba} to determine the most suitable time intervals to fit when obtaining our final masses.

We note that the principles highlighted within this section are realised only when one includes a complete set of interpolating fields effective at isolating all the states within the spectrum. This needs to include multi-particle scattering states. Our formalism is focused on the single-particle states and in particular their radial excitations. As such, Euclidean time evolution is important in suppressing contaminations from nearby scattering states. In fitting our effective mass to a plateau, we ensure single-state dominance by monitoring the $ \chi^2 $ per degree of freedom ($ \chi^2/\text{dof} $).

\section{Lattice Details}\label{sec6}
We have simulated $ \Delta^+ $ baryons on a $ 32^3\times64 $ lattice with gauge-field configurations provided by the PACS-CS Collaboration \cite{PACS-CS:2008bkb} through the ILDG \cite{Beckett:2009cb}. These configurations use an order $ a $ improved Wilson fermion action and an Iwasaki gauge action. The lattice is subject to periodic boundary conditions. The correlation functions were constructed using the COLA software library \cite{Kamleh:2022nqr}.

Our calculations are performed on gauge-field ensembles with $ \beta = 1.90 $ and a quark mass given by the hopping parameter $ \kappa = 0.13754 $, providing $ m_\pi = 413 $ MeV. At this quark mass, and with a Sommer parameter of $ r_0 = 0.4921(64) $ fm, the lattice spacing is $ a = 0.0961(13) $ fm. 

To improve the signal-to-noise ratio of our results, we perform shifts on the source insertion point and then average the resulting correlation functions. We found that at our chosen quark mass, an average over 16 source locations was sufficient.

The smearing levels used in this analysis are 16, 35, 100 and 200 sweeps. We also investigated several sets of variational parameters before settling on $ t_0 = 18 $ and $ \Delta t = 2 $ relative to a source at $ t_s = 16 $. These choices have been motivated in similar studies in the past \cite{Mahbub:2013ala}.

\section{Numerical Results}\label{sec7}
\begin{figure}[h]
	\centering
	\includegraphics[width=\linewidth]{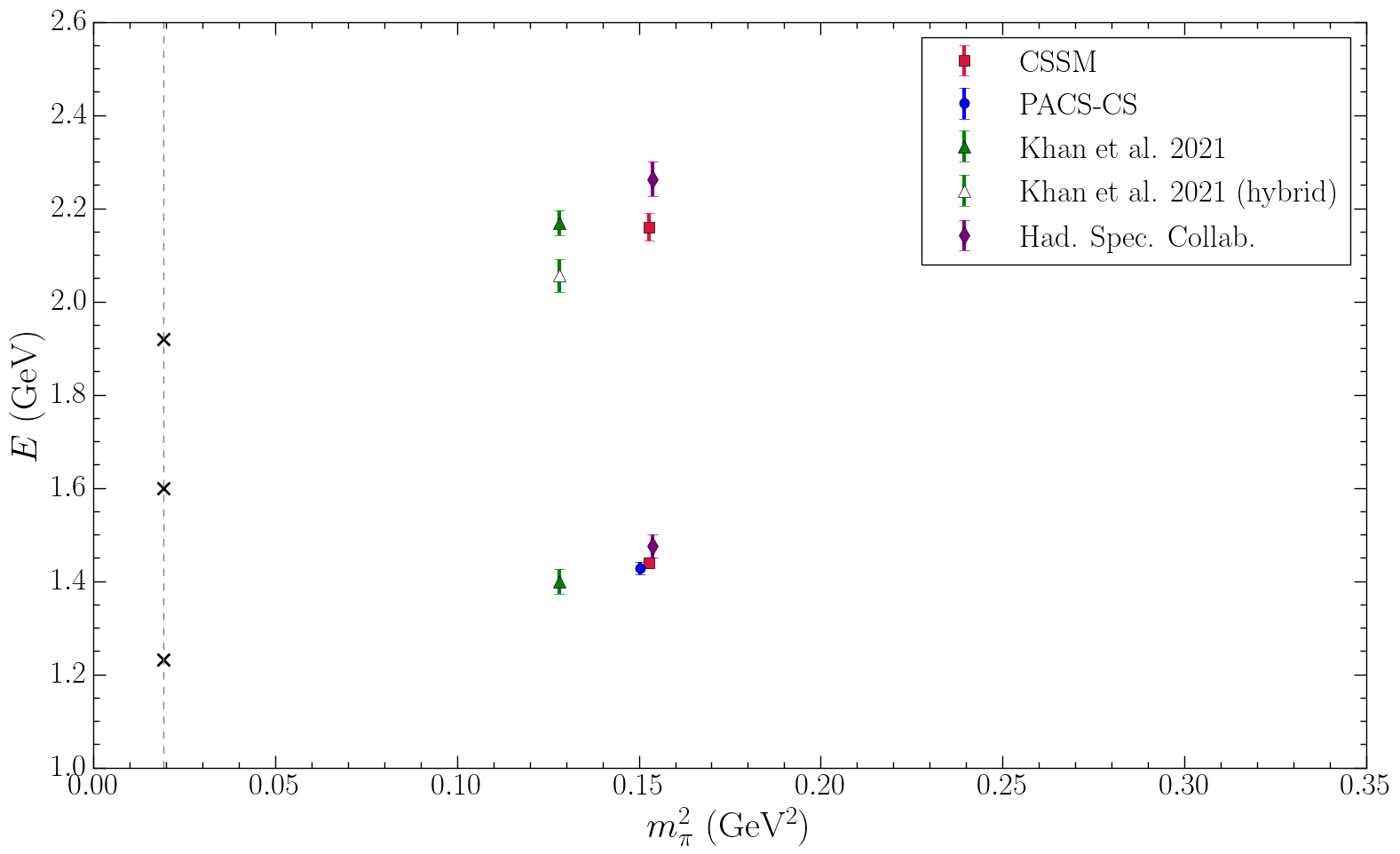}
	\caption{ Our obtained masses for the $ \Delta $ after spin-3/2 and positive parity projection (CSSM). Resonances measured experimentally are the $ \Delta(1232),\, \Delta(1600) $ and $ \Delta(1920) $, indicated by crosses along the grey dashed line at the physical point. The other studies listed here at similar quark masses to that presented here are from the PACS-CS Collaboration \cite{PACS-CS:2008bkb}, Khan et al. 2021 \cite{Khan:2020ahz} and the Hadron Spectrum Collaboration \cite{Bulava:2010yg}. Note that the PACS-CS result is offset from our own for clarity. }\label{fig2}
\end{figure}

We present results for the $ \Delta\ 3/2^+ $ spectrum in Figure~(\ref{fig2}). We have thus far been able to extract masses for the first 2 states in the spectrum and future work will aim to also extract the 2nd excited state.

Consider Figure~(\ref{fig2}), where our results are labelled `CSSM' and shown as red squares alongside other contemporary results \cite{Bulava:2010yg, Khan:2020ahz, PACS-CS:2008bkb} having a similar quark mass. While there is some indication of the ground state in our low-lying mass, there is a significant jump to get to the second state in our spectrum. This jump is more than one might expect for a state which tends to the $ \Delta(1600) $ in the physical limit. In particular, we emphasise that we don't appear to see this state using a simple 3-quark interpolating field, as given by Eq.~\eqref{deltainterp}. 

To finish this section, we have shown that a variational analysis allowed us to successfully extract $ \Delta $ masses for the ground and first excited states. We report these masses as
\begin{align}
	m_{\Delta}^{0} &= 1.439 \pm 0.007\, \text{GeV},\ \chi^2/\text{dof} = 0.972 \\
	m_{\Delta}^{1} &= 2.160 \pm 0.029\, \text{GeV},\ \chi^2/\text{dof} = 0.925
\end{align}
where the $ \chi^2/\text{dof} $ is computed via a covariance matrix analysis \cite{Mahbub:2013bba}.

\section{Eigenvector Analysis}\label{sec8}
One can also leverage the eigenvectors $ \boldsymbol{u}^{\alpha} $ obtained in our variational analysis. This has been shown useful in identifying radial excitations in the nucleon spectrum \cite{Mahbub:2013ala, Roberts:2013ipa} and we aim to apply similar techniques here. Recall that the eigenvectors were used to construct a superposition of smeared interpolating fields in Eq.~\eqref{phibar} and \eqref{phi}. We could more explicitly think of these superpositions as being written as
\begin{align}
	\overline{\phi}^\alpha 	&= u^\alpha_{16} \overline{\chi}_{16} + u^\alpha_{35} \overline{\chi}_{35} + u^\alpha_{100} \overline{\chi}_{100} + u^\alpha_{200} \overline{\chi}_{200}\,,  \\
	\phi^\alpha &= u^\alpha_{16} {\chi}_{16} + u^\alpha_{35} {\chi}_{35} + u^\alpha_{100} {\chi}_{100} + u^\alpha_{200} {\chi}_{200} \label{sm_vector}
\end{align}
where each of the $ \chi_{N_{sm}} $ is a fermion field which has been smeared $ N_{sm} $ times. As usual, the superscript $ \alpha $ is just the energy level.

Now, we already identified that smearing increases the extent of our source/sink operators. Solving the variational analysis problem is equivalent then to finding a superposition of Gaussians which has largest overlap with the wave function of the desired $ \Delta $ state.

This simple reinterpretation of the superpositions gives us a powerful tool for classifying the states in the spectrum. Consider the case of superposing, for example, a narrow Gaussian with positive sign, with a wider Gaussian of negative signature. The resulting distribution will have a crossing through zero at the edge of the positive Gaussian, where the superposition becomes dominated by the negatively weighted Gaussian. In other words, if there is a relative sign between consecutive Gaussians in our superposition, this will indicate a zero crossing or a node in the radial part of the wave function. 

\begin{figure}[h]
	\centering
	\includegraphics[width=\linewidth]{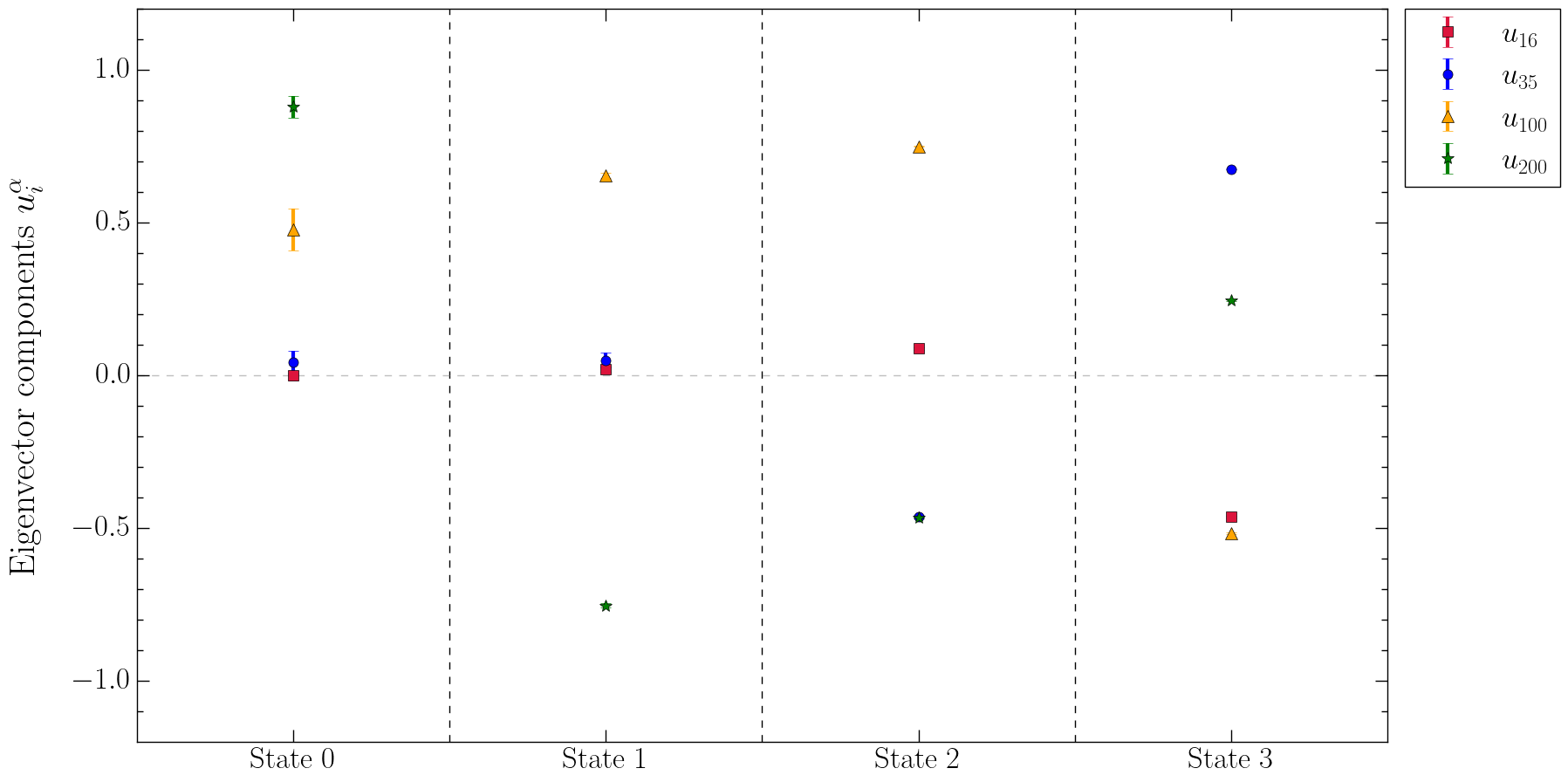}
	\caption{ Components of the normalised eigenvectors $ u^\alpha_i $. The figure is composed of four subplots with each labelled along the x-axis by state $ \alpha $. The components of the eigenvectors are labelled by the number of smearing sweeps, indicated in the legend. }\label{fig3}
\end{figure}

Consider Figure~(\ref{fig3}). This illustrates the superposition of Gaussians via the $ u^\alpha_i $ components obtained through the variational method. More specifically, let us focus on the subplot for State 0. All components of the eignvector are positive or approximately zero, so we would be inclined to think there is little evidence for a zero crossing. We would thus identify this state as a simple $ 1s $ state, since it has no nodes.

Moving onto the State 1 plot, we see that both the 16 and 35 smearing components are centred about zero. The 100 component is clearly positive and large, and the 200 component is negative and large. We would interpret this exactly as the toy case above, where there is a narrow peak of positive signature superposed with a broader peak of negative signature. Once the positive peak becomes less prominent, the negative Gaussian will take over and there will be 1 zero crossing. We would identify this as a $ 2s $ state as the superposition has a single node. Similar analysis of the State 2 and 3 subplots shows 2 and 3 nodes respectively, so we would identify these as $ 3s $ and $ 4s $ states respectively.

A more involved analysis can be performed whereby one computes the wave function of one of the light quarks within the baryon directly using lattice QCD \cite{Roberts:2013ipa}. Such a computation would be a useful point of comparison, but leveraging the variational method eigenvectors as discussed above offers an alternative for classifying states in the spectrum.

\section{Hamiltonian Effective Field Theory}\label{sec9}
Looking ahead to future studies, we briefly allude to Hamiltonian Effective Field Theory (HEFT) and its use in studying low-lying baryon resonances \cite{Wu:2017qve, Abell:2021awi}. Essentially, HEFT bridges the gap between the finite volume of the lattice, and what is observed in the infinite volume of the real world (at particle collider experiments for example). By considering a Hamiltonian composed of single-particle bare states and meson-baryon multi-particle states, one is able to construct a scattering model which can be fit to experimental observables, such as phase shifts and inelasticities in, say, $ \pi N $ scattering.

Such a model can then be brought to the finite volume of lattice QCD and extrapolated to unphysical quark masses to confront lattice QCD data, such as that obtained in Figure~(\ref{fig2}). Having a broad quark mass range to perform fits to affords a good opportunity for such comparisons. HEFT has been shown to provide new understanding for the nature of the Roper resonance \cite{Wu:2017qve, Liu:2016uzk}, the $ \Lambda(1405) $ \cite{Liu:2016wxq, Hall:2014uca} and for the low-lying $ \Delta $ spectrum \cite{Abell:2021awi, Hall:2013qba}. We aim to extend this work using the updated lattice QCD masses reported in this paper. 

\section{Conclusion}\label{sec10}
In this preliminary investigation into the $ \Delta\ 3/2^+ $ spectrum, we have obtained ground state and first excited state finite-volume masses at a quark mass corresponding to $ m_\pi~=~413 $ MeV. Our ground state results are comparable with the earlier PACS-CS results, and our 1st excitation is similar to other recent analyses, sitting higher than the expected $ \Delta(1600) $ state. 

Building on previous studies using smeared interpolating fields, we have been able to identify both states in our spectrum by their nodal structure as being a $ 1s $ state and $ 2s $ state. We hope to take this work and make connection to experimental data through HEFT and thus glean some deeper understanding of resonance structure for the low energy $ \Delta $ excitations.

\backmatter

%

\bmhead{Acknowledgements}
L.H. would like to thank Ryan Bignell for providing code and advice in the early stages of this work. 

\section*{Statements and Declarations}
\subsection*{Funding}
This work was supported by the Australian Government Research Training Program Scholarship and with supercomputing resources provided by the Phoenix HPC service at the University of Adelaide. This research was undertaken with the assistance of resources from the National Computational Infrastructure (NCI), provided through the National Computational Merit Allocation Scheme, and supported by the Australian Government through Grant No. LE190100021 via the University of Adelaide Partner Share. This research was supported by the Australian Research Council through Grants No. DP190102215 and DP210103706. W.K. was supported by the PAWSEY Supercomputing Centre through the Pawsey Centre for Extreme Scale Readiness (PaCER) program.

\subsection*{Competing Interests}
The authors have no competing interests to declare that are relevant to the content of this article.

\bibliography{sn-bibliography}


\begin{thebibliography}{20}
\ifx \bisbn   \undefined \def \bisbn  #1{ISBN #1}\fi
\ifx \binits  \undefined \def \binits#1{#1}\fi
\ifx \bauthor  \undefined \def \bauthor#1{#1}\fi
\ifx \batitle  \undefined \def \batitle#1{#1}\fi
\ifx \bjtitle  \undefined \def \bjtitle#1{#1}\fi
\ifx \bvolume  \undefined \def \bvolume#1{\textbf{#1}}\fi
\ifx \byear  \undefined \def \byear#1{#1}\fi
\ifx \bissue  \undefined \def \bissue#1{#1}\fi
\ifx \bfpage  \undefined \def \bfpage#1{#1}\fi
\ifx \blpage  \undefined \def \blpage #1{#1}\fi
\ifx \burl  \undefined \def \burl#1{\textsf{#1}}\fi
\ifx \doiurl  \undefined \def \doiurl#1{\url{https://doi.org/#1}}\fi
\ifx \betal  \undefined \def \betal{\textit{et al.}}\fi
\ifx \binstitute  \undefined \def \binstitute#1{#1}\fi
\ifx \binstitutionaled  \undefined \def \binstitutionaled#1{#1}\fi
\ifx \bctitle  \undefined \def \bctitle#1{#1}\fi
\ifx \beditor  \undefined \def \beditor#1{#1}\fi
\ifx \bpublisher  \undefined \def \bpublisher#1{#1}\fi
\ifx \bbtitle  \undefined \def \bbtitle#1{#1}\fi
\ifx \bedition  \undefined \def \bedition#1{#1}\fi
\ifx \bseriesno  \undefined \def \bseriesno#1{#1}\fi
\ifx \blocation  \undefined \def \blocation#1{#1}\fi
\ifx \bsertitle  \undefined \def \bsertitle#1{#1}\fi
\ifx \bsnm \undefined \def \bsnm#1{#1}\fi
\ifx \bsuffix \undefined \def \bsuffix#1{#1}\fi
\ifx \bparticle \undefined \def \bparticle#1{#1}\fi
\ifx \barticle \undefined \def \barticle#1{#1}\fi
\bibcommenthead
\ifx \bconfdate \undefined \def \bconfdate #1{#1}\fi
\ifx \botherref \undefined \def \botherref #1{#1}\fi
\ifx \url \undefined \def \url#1{\textsf{#1}}\fi
\ifx \bchapter \undefined \def \bchapter#1{#1}\fi
\ifx \bbook \undefined \def \bbook#1{#1}\fi
\ifx \bcomment \undefined \def \bcomment#1{#1}\fi
\ifx \oauthor \undefined \def \oauthor#1{#1}\fi
\ifx \citeauthoryear \undefined \def \citeauthoryear#1{#1}\fi
\ifx \endbibitem  \undefined \def \endbibitem {}\fi
\ifx \bconflocation  \undefined \def \bconflocation#1{#1}\fi
\ifx \arxivurl  \undefined \def \arxivurl#1{\textsf{#1}}\fi
\csname PreBibitemsHook\endcsname

\bibitem[\protect\citeauthoryear{Burkert and Roberts}{2019}]{Burkert:2017djo}
\begin{barticle}
\bauthor{\bsnm{Burkert}, \binits{V.D.}},
\bauthor{\bsnm{Roberts}, \binits{C.D.}}:
\batitle{{Colloquium : Roper resonance: Toward a solution to the fifty year
  puzzle}}.
\bjtitle{Rev. Mod. Phys.}
\bvolume{91}(\bissue{1}),
\bfpage{011003}
(\byear{2019})
\doiurl{10.1103/RevModPhys.91.011003}
{\href{https://arxiv.org/abs/1710.02549}{{arXiv:1710.02549}}}
{[nucl-ex]}
\end{barticle}
\endbibitem

\bibitem[\protect\citeauthoryear{Wu et~al.}{2018}]{Wu:2017qve}
\begin{barticle}
\bauthor{\bsnm{Wu}, \binits{J.-j.}},
\bauthor{\bsnm{Leinweber}, \binits{D.B.}},
\bauthor{\bsnm{Liu}, \binits{Z.-w.}},
\bauthor{\bsnm{Thomas}, \binits{A.W.}}:
\batitle{{Structure of the Roper Resonance from Lattice QCD Constraints}}.
\bjtitle{Phys. Rev. D}
\bvolume{97}(\bissue{9}),
\bfpage{094509}
(\byear{2018})
\doiurl{10.1103/PhysRevD.97.094509}
{\href{https://arxiv.org/abs/1703.10715}{{arXiv:1703.10715}}}
{[nucl-th]}
\end{barticle}
\endbibitem

\bibitem[\protect\citeauthoryear{Hall et~al.}{2013}]{Hall:2013qba}
\begin{barticle}
\bauthor{\bsnm{Hall}, \binits{J.M.M.}},
\bauthor{\bsnm{Hsu}, \binits{A.C.-P.}},
\bauthor{\bsnm{Leinweber}, \binits{D.B.}},
\bauthor{\bsnm{Thomas}, \binits{A.W.}},
\bauthor{\bsnm{Young}, \binits{R.D.}}:
\batitle{{Finite-volume matrix Hamiltonian model for a $\Delta \to N\pi$
  system}}.
\bjtitle{Phys. Rev. D}
\bvolume{87}(\bissue{9}),
\bfpage{094510}
(\byear{2013})
\doiurl{10.1103/PhysRevD.87.094510}
{\href{https://arxiv.org/abs/1303.4157}{{arXiv:1303.4157}}}
{[hep-lat]}
\end{barticle}
\endbibitem

\bibitem[\protect\citeauthoryear{Hall et~al.}{2015}]{Hall:2014uca}
\begin{barticle}
\bauthor{\bsnm{Hall}, \binits{J.M.M.}},
\bauthor{\bsnm{Kamleh}, \binits{W.}},
\bauthor{\bsnm{Leinweber}, \binits{D.B.}},
\bauthor{\bsnm{Menadue}, \binits{B.J.}},
\bauthor{\bsnm{Owen}, \binits{B.J.}},
\bauthor{\bsnm{Thomas}, \binits{A.W.}},
\bauthor{\bsnm{Young}, \binits{R.D.}}:
\batitle{{Lattice QCD Evidence that the \ensuremath{\Lambda}(1405) Resonance is
  an Antikaon-Nucleon Molecule}}.
\bjtitle{Phys. Rev. Lett.}
\bvolume{114}(\bissue{13}),
\bfpage{132002}
(\byear{2015})
\doiurl{10.1103/PhysRevLett.114.132002}
{\href{https://arxiv.org/abs/1411.3402}{{arXiv:1411.3402}}}
{[hep-lat]}
\end{barticle}
\endbibitem

\bibitem[\protect\citeauthoryear{Liu et~al.}{2016}]{Liu:2015ktc}
\begin{barticle}
\bauthor{\bsnm{Liu}, \binits{Z.-W.}},
\bauthor{\bsnm{Kamleh}, \binits{W.}},
\bauthor{\bsnm{Leinweber}, \binits{D.B.}},
\bauthor{\bsnm{Stokes}, \binits{F.M.}},
\bauthor{\bsnm{Thomas}, \binits{A.W.}},
\bauthor{\bsnm{Wu}, \binits{J.-J.}}:
\batitle{{Hamiltonian effective field theory study of the $\mathbf{N^*(1535)}$
  resonance in lattice QCD}}.
\bjtitle{Phys. Rev. Lett.}
\bvolume{116}(\bissue{8}),
\bfpage{082004}
(\byear{2016})
\doiurl{10.1103/PhysRevLett.116.082004}
{\href{https://arxiv.org/abs/1512.00140}{{arXiv:1512.00140}}}
{[hep-lat]}
\end{barticle}
\endbibitem

\bibitem[\protect\citeauthoryear{Liu et~al.}{2017a}]{Liu:2016uzk}
\begin{barticle}
\bauthor{\bsnm{Liu}, \binits{Z.-W.}},
\bauthor{\bsnm{Kamleh}, \binits{W.}},
\bauthor{\bsnm{Leinweber}, \binits{D.B.}},
\bauthor{\bsnm{Stokes}, \binits{F.M.}},
\bauthor{\bsnm{Thomas}, \binits{A.W.}},
\bauthor{\bsnm{Wu}, \binits{J.-J.}}:
\batitle{{Hamiltonian effective field theory study of the $\mathbf{N^*(1440)}$
  resonance in lattice QCD}}.
\bjtitle{Phys. Rev. D}
\bvolume{95}(\bissue{3}),
\bfpage{034034}
(\byear{2017})
\doiurl{10.1103/PhysRevD.95.034034}
{\href{https://arxiv.org/abs/1607.04536}{{arXiv:1607.04536}}}
{[nucl-th]}
\end{barticle}
\endbibitem

\bibitem[\protect\citeauthoryear{Liu et~al.}{2017b}]{Liu:2016wxq}
\begin{barticle}
\bauthor{\bsnm{Liu}, \binits{Z.-W.}},
\bauthor{\bsnm{Hall}, \binits{J.M.M.}},
\bauthor{\bsnm{Leinweber}, \binits{D.B.}},
\bauthor{\bsnm{Thomas}, \binits{A.W.}},
\bauthor{\bsnm{Wu}, \binits{J.-J.}}:
\batitle{{Structure of the $\mathbf{\Lambda(1405)}$ from Hamiltonian effective
  field theory}}.
\bjtitle{Phys. Rev. D}
\bvolume{95}(\bissue{1}),
\bfpage{014506}
(\byear{2017})
\doiurl{10.1103/PhysRevD.95.014506}
{\href{https://arxiv.org/abs/1607.05856}{{arXiv:1607.05856}}}
{[nucl-th]}
\end{barticle}
\endbibitem

\bibitem[\protect\citeauthoryear{Abell et~al.}{2022}]{Abell:2021awi}
\begin{barticle}
\bauthor{\bsnm{Abell}, \binits{C.D.}},
\bauthor{\bsnm{Leinweber}, \binits{D.B.}},
\bauthor{\bsnm{Thomas}, \binits{A.W.}},
\bauthor{\bsnm{Wu}, \binits{J.-J.}}:
\batitle{{Regularization in nonperturbative extensions of effective field
  theory}}.
\bjtitle{Phys. Rev. D}
\bvolume{106}(\bissue{3}),
\bfpage{034506}
(\byear{2022})
\doiurl{10.1103/PhysRevD.106.034506}
{\href{https://arxiv.org/abs/2110.14113}{{arXiv:2110.14113}}}
{[hep-lat]}
\end{barticle}
\endbibitem

\bibitem[\protect\citeauthoryear{Mahbub et~al.}{2013}]{Mahbub:2013ala}
\begin{barticle}
\bauthor{\bsnm{Mahbub}, \binits{M.S.}},
\bauthor{\bsnm{Kamleh}, \binits{W.}},
\bauthor{\bsnm{Leinweber}, \binits{D.B.}},
\bauthor{\bsnm{Moran}, \binits{P.J.}},
\bauthor{\bsnm{Williams}, \binits{A.G.}}:
\batitle{{Structure and Flow of the Nucleon Eigenstates in Lattice QCD}}.
\bjtitle{Phys. Rev. D}
\bvolume{87}(\bissue{9}),
\bfpage{094506}
(\byear{2013})
\doiurl{10.1103/PhysRevD.87.094506}
{\href{https://arxiv.org/abs/1302.2987}{{arXiv:1302.2987}}}
{[hep-lat]}
\end{barticle}
\endbibitem

\bibitem[\protect\citeauthoryear{Zanotti et~al.}{2003}]{Zanotti:2003fx}
\begin{barticle}
\bauthor{\bsnm{Zanotti}, \binits{J.M.}},
\bauthor{\bsnm{Leinweber}, \binits{D.B.}},
\bauthor{\bsnm{Williams}, \binits{A.G.}},
\bauthor{\bsnm{Zhang}, \binits{J.B.}},
\bauthor{\bsnm{Melnitchouk}, \binits{W.}},
\bauthor{\bsnm{Choe}, \binits{S.}}:
\batitle{{Spin 3/2 nucleon and delta baryons in lattice QCD}}.
\bjtitle{Phys. Rev. D}
\bvolume{68},
\bfpage{054506}
(\byear{2003})
\doiurl{10.1103/PhysRevD.68.054506}
{\href{https://arxiv.org/abs/hep-lat/0304001}{{arXiv:hep-lat/0304001}}}
\end{barticle}
\endbibitem

\bibitem[\protect\citeauthoryear{Leinweber et~al.}{1992}]{Leinweber:1992hy}
\begin{barticle}
\bauthor{\bsnm{Leinweber}, \binits{D.B.}},
\bauthor{\bsnm{Draper}, \binits{T.}},
\bauthor{\bsnm{Woloshyn}, \binits{R.M.}}:
\batitle{{Decuplet baryon structure from lattice QCD}}.
\bjtitle{Phys. Rev. D}
\bvolume{46},
\bfpage{3067}--\blpage{3085}
(\byear{1992})
\doiurl{10.1103/PhysRevD.46.3067}
{\href{https://arxiv.org/abs/hep-lat/9208025}{{arXiv:hep-lat/9208025}}}
\end{barticle}
\endbibitem

\bibitem[\protect\citeauthoryear{Benmerrouche
  et~al.}{1989}]{Benmerrouche:1989uc}
\begin{barticle}
\bauthor{\bsnm{Benmerrouche}, \binits{M.}},
\bauthor{\bsnm{Davidson}, \binits{R.M.}},
\bauthor{\bsnm{Mukhopadhyay}, \binits{N.C.}}:
\batitle{Problems of describing spin 3/2 baryon resonances in the effective
  lagrangian theory}.
\bjtitle{Phys. Rev. C}
\bvolume{39},
\bfpage{2339}--\blpage{2348}
(\byear{1989})
\doiurl{10.1103/PhysRevC.39.2339}
\end{barticle}
\endbibitem

\bibitem[\protect\citeauthoryear{Michael}{1985}]{Michael:1985ne}
\begin{barticle}
\bauthor{\bsnm{Michael}, \binits{C.}}:
\batitle{{Adjoint Sources in Lattice Gauge Theory}}.
\bjtitle{Nucl. Phys. B}
\bvolume{259},
\bfpage{58}--\blpage{76}
(\byear{1985})
\doiurl{10.1016/0550-3213(85)90297-4}
\end{barticle}
\endbibitem

\bibitem[\protect\citeauthoryear{Mahbub et~al.}{2014}]{Mahbub:2013bba}
\begin{barticle}
\bauthor{\bsnm{Mahbub}, \binits{M.S.}},
\bauthor{\bsnm{Kamleh}, \binits{W.}},
\bauthor{\bsnm{Leinweber}, \binits{D.B.}},
\bauthor{\bsnm{Williams}, \binits{A.G.}}:
\batitle{{Searching for low-lying multi-particle thresholds in lattice
  spectroscopy}}.
\bjtitle{Annals Phys.}
\bvolume{342},
\bfpage{270}--\blpage{282}
(\byear{2014})
\doiurl{10.1016/j.aop.2014.01.004}
{\href{https://arxiv.org/abs/1310.6803}{{arXiv:1310.6803}}}
{[hep-lat]}
\end{barticle}
\endbibitem

\bibitem[\protect\citeauthoryear{Aoki et~al.}{2009}]{PACS-CS:2008bkb}
\begin{barticle}
\bauthor{\bsnm{Aoki}, \binits{S.}}, \betal:
\batitle{{2+1 Flavor Lattice QCD toward the Physical Point}}.
\bjtitle{Phys. Rev. D}
\bvolume{79},
\bfpage{034503}
(\byear{2009})
\doiurl{10.1103/PhysRevD.79.034503}
{\href{https://arxiv.org/abs/0807.1661}{{arXiv:0807.1661}}}
{[hep-lat]}
\end{barticle}
\endbibitem

\bibitem[\protect\citeauthoryear{Beckett et~al.}{2011}]{Beckett:2009cb}
\begin{barticle}
\bauthor{\bsnm{Beckett}, \binits{M.G.}},
\bauthor{\bsnm{Joo}, \binits{B.}},
\bauthor{\bsnm{Maynard}, \binits{C.M.}},
\bauthor{\bsnm{Pleiter}, \binits{D.}},
\bauthor{\bsnm{Tatebe}, \binits{O.}},
\bauthor{\bsnm{Yoshie}, \binits{T.}}:
\batitle{{Building the International Lattice Data Grid}}.
\bjtitle{Comput. Phys. Commun.}
\bvolume{182},
\bfpage{1208}--\blpage{1214}
(\byear{2011})
\doiurl{10.1016/j.cpc.2011.01.027}
{\href{https://arxiv.org/abs/0910.1692}{{arXiv:0910.1692}}}
{[hep-lat]}
\end{barticle}
\endbibitem

\bibitem[\protect\citeauthoryear{Kamleh}{2023}]{Kamleh:2022nqr}
\begin{barticle}
\bauthor{\bsnm{Kamleh}, \binits{W.}}:
\batitle{{Evolving the COLA software library}}.
\bjtitle{PoS}
\bvolume{LATTICE2022},
\bfpage{339}
(\byear{2023})
\doiurl{10.22323/1.430.0339}
{\href{https://arxiv.org/abs/2302.00850}{{arXiv:2302.00850}}}
{[hep-lat]}
\end{barticle}
\endbibitem

\bibitem[\protect\citeauthoryear{Khan et~al.}{2021}]{Khan:2020ahz}
\begin{barticle}
\bauthor{\bsnm{Khan}, \binits{T.}},
\bauthor{\bsnm{Richards}, \binits{D.}},
\bauthor{\bsnm{Winter}, \binits{F.}}:
\batitle{{Positive-parity baryon spectrum and the role of hybrid baryons}}.
\bjtitle{Phys. Rev. D}
\bvolume{104}(\bissue{3}),
\bfpage{034503}
(\byear{2021})
\doiurl{10.1103/PhysRevD.104.034503}
{\href{https://arxiv.org/abs/2010.03052}{{arXiv:2010.03052}}}
{[hep-lat]}
\end{barticle}
\endbibitem

\bibitem[\protect\citeauthoryear{Bulava et~al.}{2010}]{Bulava:2010yg}
\begin{barticle}
\bauthor{\bsnm{Bulava}, \binits{J.}},
\bauthor{\bsnm{Edwards}, \binits{R.G.}},
\bauthor{\bsnm{Engelson}, \binits{E.}},
\bauthor{\bsnm{Joo}, \binits{B.}},
\bauthor{\bsnm{Lin}, \binits{H.-W.}},
\bauthor{\bsnm{Morningstar}, \binits{C.}},
\bauthor{\bsnm{Richards}, \binits{D.G.}},
\bauthor{\bsnm{Wallace}, \binits{S.J.}}:
\batitle{{Nucleon, $\Delta$ and $\Omega$ excited states in $N_f=2+1$ lattice
  QCD}}.
\bjtitle{Phys. Rev. D}
\bvolume{82},
\bfpage{014507}
(\byear{2010})
\doiurl{10.1103/PhysRevD.82.014507}
{\href{https://arxiv.org/abs/1004.5072}{{arXiv:1004.5072}}}
{[hep-lat]}
\end{barticle}
\endbibitem

\bibitem[\protect\citeauthoryear{Roberts et~al.}{2013}]{Roberts:2013ipa}
\begin{barticle}
\bauthor{\bsnm{Roberts}, \binits{D.S.}},
\bauthor{\bsnm{Kamleh}, \binits{W.}},
\bauthor{\bsnm{Leinweber}, \binits{D.B.}}:
\batitle{{Wave Function of the Roper from Lattice QCD}}.
\bjtitle{Phys. Lett. B}
\bvolume{725},
\bfpage{164}--\blpage{169}
(\byear{2013})
\doiurl{10.1016/j.physletb.2013.06.056}
{\href{https://arxiv.org/abs/1304.0325}{{arXiv:1304.0325}}}
{[hep-lat]}
\end{barticle}
\endbibitem

\end{thebibliography}

\end{document}